\documentclass[prl,twocolumn,superscriptaddress,showpacs,floatfix,nofootinbib]{revtex4-1}
\usepackage[english]{babel}
\usepackage[dvips]{graphicx}
\usepackage{amsmath}
\usepackage{amsfonts}
\usepackage{amssymb}
\usepackage{verbatim}

\begin{document}
\title{Cooling of nanomechanical resonator by thermally activated single-electron transport}

\author{F. Santandrea}
\affiliation{Department of Physics, University of Gothenburg, SE - 412 96
G{\"o}teborg, Sweden}
\author{L. Y. Gorelik}
\affiliation{Department of Applied Physics, Chalmers University of Technology,
SE - 412 96 G{\"o}teborg, Sweden}
\author{R. I. Shekhter}
\affiliation{Department of Physics, University of Gothenburg, SE - 412 96
G{\"o}teborg, Sweden}
\author{M. Jonson}
\affiliation{Department of Physics, University of Gothenburg, SE -
412 96 G{\"o}teborg, Sweden}
\affiliation{SUPA, Department of Physics, Heriot-Watt University,
Edinburgh EH14 4AS, Scotland, UK}
\affiliation{Division of Quantum Phases and Devices, School of Physics,
Konkuk University, Seoul 143-107, Korea}

\begin{abstract}
We show that the vibrations of a nanomechanical resonator can be cooled 
to near its quantum ground state by tunnelling injection of electrons from an 
STM tip.  The interplay between two mechanisms for coupling the electronic 
and mechanical degrees of freedom results in a bias-voltage dependent 
difference between the probability amplitudes for vibron emission and 
absorption during tunneling. For a bias voltage just below the Coulomb 
blockade threshold we find that absorption dominates, which leads to cooling 
corresponding to an average vibron population of the fundamental bending 
mode of 0.2. 
\end{abstract}


\maketitle


\noindent
Remarkable steps are now being taken towards achieving conditions under 
which quantum effects are experimentally accessible in nano-electromechanical 
systems (NEMS)\cite{O'Connell2010}. This is encouraging both for the prospects 
of realizing a plethora of applications that depend on our ability to control and 
monitor the coherent dynamics of nanometer-scale mechanical systems and for 
shedding light on purely fundamental issues, such as the nature of the crossover 
from classical to quantum physics \cite{Blencowe2004}.

A number of proposals have been put forward in order to reach the extremely low 
temperatures $T$ of order $\hbar\omega/k_B$,  where quantum effects become observable 
in mechanical resonators of eigenfrequency $\omega$. Their aim has been to replace the
conventional dilution refrigerators by more efficient active cooling methods.

Most of them are based on well-established principles for laser cooling of atoms and 
molecules \cite{Wilson-Rae2004}, but alternative approaches have
been proposed by several authors who suggest that the coupling 
between mechanical and electronic degrees of freedom can be exploited for inducing 
energy to flow from the former to the latter  \cite{Zippilli2009,*Pistolesi2009,*Ouyang2009,*Sonne2010}.
All these schemes for ground state cooling are based on quantum resolved sideband 
transitions between discrete quantum levels of the refrigerant, a cooling mechanism 
that relies on \textit{energy conservation} in order to suppress 
processes which involve emission of vibrational energy quanta (vibrons) with 
respect to those that lead to absorption of such quanta.

In this Letter we suggest a new mechanism for ground state cooling of a nanomechanical 
resonator, which is based on passing a current through the resonator under conditions 
such that the \textit{probability amplitude} for tunneling electrons to emit vibrons is much 
lower than it is to absorb them. As such it has the advantage that it does not require the 
refrigerant to have a discrete energy spectrum, which puts fewer constraints on the 
experimental design.

\begin{figure}
\center
\includegraphics[width=0.4\textwidth]{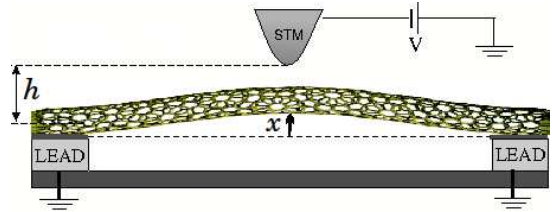}
\caption{Sketch of the model system considered. A metallic carbon nanotube is 
suspended over a trench between two grounded metallic leads, while an STM tip 
placed a distance $h$ above the nanotube is biased at a negative voltage $-V$.}
\label{fig:system}
\end{figure}

To be specific we consider the model system sketched in Fig.~\ref{fig:system}, where 
electrons are injected from the tip of a scanning tunneling microscope (STM) into a
suspended metallic carbon nanotube. Low-temperature tunneling spectroscopy studies 
on a similar device \cite{LeRoy2004} have shown that inelastic electron tunneling 
can create a non-thermal equilibrium population of vibronic states in 
the nanotube. Below, we will show that the probability for vibron emission can 
be suppressed as a result of the interference between two different mechanisms for
coupling the mechanical and electronic degrees of freedom of the system. One of these 
mechanisms is the nanotube-position dependent probability amplitude for electron 
tunneling from the STM tip to the nanotube, the other is the electrostatic force on the 
nanotube when it is charged.  

It turns out that the effect of the interference depends on the voltage bias between the 
STM tip and the leads. Our analysis shows that the destructive interference is maximal
for a bias voltage slightly below the threshold voltage for lifting the Coulomb blockade 
of electron tunneling through the system. If the nanotube is weakly enough coupled to 
the environment, the suppression of vibron emission is strong enough to drive the 
nanotube to near its vibrational ground state and hence effectively ``cool'' the 
mechanical degrees of  freedom.

In order to analyze the dynamics of the nanotube and of the tunneling electrons
in the quantum regime we introduce a model Hamiltonian,
\begin{equation}\label{eq:H_tot} 
H =  H_e+H_m+H_t+H_C, 
\end{equation}
\noindent where
\begin{align}
H_e & =\sum_{q,\alpha} E_{q,\alpha} a_{q,\alpha}^{\dagger} a_{q,\alpha}+\sum_q \xi_q c^\dagger_q c_q \,,\\
H_m & =\hbar\omega \left( b^{\dag}b +1/2\right) \,,\\
H_t & =\sum_{q,q'}e^{i\hat{\varphi}} c_{q'}^\dagger\left[ t_S(\hat{X}) a_{q,S} + t_{L} a_{q,L}\right]+h.c.,
\label{eq:H_t}
\end{align}
\noindent and where $a_{q,\alpha}^{(\dagger)}$ and $c_q^{(\dagger)}$ are annihilation 
(creation) operators for electrons in the STM tip ($\alpha=S$), in the leads 
($\alpha=L$), and in the nanotube, respectively.

The first term of (\ref{eq:H_tot}), $H_e$, describes the STM tip, 
the leads, and the nanotube as reservoirs of non-interacting electrons. The 
second term, $H_m$, describes  the nanotube's mechanical degrees of 
freedom, which we restrict to the fundamental bending mode considered as a 
simple harmonic oscillator with angular frequency $\omega$, $b^{(\dagger)}$ 
being the annihilation (creation) operator for an elementary excitation (vibron) 
of this mode.

Electron tunneling between the STM tip and the nanotube and between the 
nanotube and the leads are described by $H_t$, the third term of the Hamiltonian,
in terms of the tunneling amplitudes  $t_S$ and  $t_L$. Here the operator 
$e^{i\hat{\varphi}}$ changes the number $N$ of excess electrons on the nanotube by 
one,  $e^{-i\hat{\varphi}}\hat{N}e^{i\hat{\varphi}} = \hat{N} + 1$. Since $t_S$ depends on 
the overlap between electronic states in the STM tip and the nanotube,  it depends
on the deflection of the tube through the operator $\hat{X}=\Delta x_{gs}(b^{\dagger}+b)$, 
where $\Delta x_{gs} \equiv \sqrt{\hbar/(2M\omega)}$ is the 
displacement uncertainty in the vibrational ground state and $M$ is an effective 
oscillator mass. For simplicity we assume that the STM tip is positioned above 
the midpoint of the nanotube (see Fig.~\ref{fig:system}) and model the deflection 
dependence of the tunneling amplitude as $t_S(\hat{X}) \equiv t_S\exp(\hat{X}/\lambda)$, 
where $\lambda$ is the characteristic tunneling length of the barrier ($\lambda \simeq 10^{-10}$ m). 
This dependence amounts to a coupling between the electronic and mechanical degrees of 
freedom that we will refer to as a \textit{tunneling} electromechanical (TEM) 
coupling. In contrast, the distance between the nanotube and the leads is fixed, 
so that $t_L$ does not depend on the nanotube deflection.

The last term, $H_C$, in (\ref{eq:H_tot}) describes the electrostatic interactions 
in the system, which we will treat in the framework of the capacitance model.
In this approximation $H_C$ only depends on the total charge on the 
nanotube and on the voltages applied to the bulk electrodes. Assuming the 
supporting leads to be grounded and that a negative 
electrostatic potential $-V$ ($V>0$) is applied to the STM electrode, 
we restrict our analysis to the Coulomb blockade regime in which at most 
one extra electron may reside on the nanotube. Under such conditions 
$H_C$ can be written as \cite{Landau1998}
\begin{equation}\label{eq:H_C}
H_C=e\left[\frac{C_g(V_C-V)}{C_{\Sigma}}+V\right] \hat{N}
-\frac{C_SC_gV^2}{2C_\Sigma}\,,
\end{equation}
where $C_\Sigma=C_S+C_g$, $C_S$ is the mutual 
capacitance between nanotube and STM tip, $C_g$ is the total capacitance 
between nanotube and ground, $V_C=e/2C_g$ is the threshold value of $V$
for lifting the Coulomb blockade and $-e$ is the electronic charge. 

In general, $C_S$ and $ C_g$ both depend on the geometry of the system 
and therefore on the nanotube deflection. Here we will only take 
the dominant deflection dependance of the STM-nanotube capacitance into 
account. Hence $C_S=C_S(h-\hat{X})$, where $h$ is the distance between 
the STM and the straight nanotube. For small displacements of the nanotube 
we may linearize the interaction Hamiltonian (\ref{eq:H_C}) and use an 
approximation that for $C_S \ll C_g$ takes the form
%
%
\begin{equation}\label{eq:H_C-linear}
H_C = U_C(V)\hat{N} -\mathfrak{F}\hat{X}\hat{N}-\alpha(\hat{X})V^2,
\end{equation}

\noindent where $\mathfrak{F} \equiv 2(\partial C_S /\partial x)_0 V_C\delta V$,
and $\delta V \equiv V_C-V$. The first term of (\ref{eq:H_C-linear}) 
determines the Coulomb blockade effect  in the absence of nanotube deflections, 
while the second is a deflection-dependent electromechanical interaction term. 
Due to a formal analogy with the interaction term in the model Hamiltonian 
for the polaron problem, we will refer to the origin of this term as a 
\textit{polaronic} electromechanical (PEM) coupling. The last term of 
(\ref{eq:H_C-linear}) is a contribution that does not 
depend on whether the nanotube is charged or not. 

It is important for what follows that the sign of the polaronic force constant  
$\mathfrak{F}$ in (\ref{eq:H_C-linear}) depends on the bias voltage. If the bias 
voltage is below the Coulomb blockade threshold, so that only thermally activated 
transport is possible, i.e. if  $\delta V>0$, then $\mathfrak{F}>0$ and hence if 
charged by an electron the nanotube will be attracted to the STM tip. On the 
other hand, if $\delta V<0$, then $\mathfrak{F}<0$ and the 
charged nanotube is repelled from the STM. 

Note that the possibility to change the direction of the force by varying the bias voltage crucially 
relies on the discrete nature of the tunneling charge.  If this charge could be 
arbitrarily small, then $V_C \to 0$ and hence $\mathfrak{F}\propto -V$.  For any 
(positive) value of $V$ the polaronic force would therefore be negative and 
push the charged nanotube away from the STM tip, decreasing the tunneling 
matrix element \cite{Jonsson2005}.

As we have seen above, the electromechanical interaction is described by 
two separate terms in the Hamiltonian, one due to what we call TEM coupling 
and the other due to PEM coupling. The cooling mechanism to be discussed 
below results from the interplay between these different types of coupling. In 
order to analyze this interplay, it is convenient to apply a unitary transformation 
that removes the polaronic term from the Hamiltonian and instead makes the 
tunneling amplitudes dependent on both the midpoint position $\hat{X}$ of the 
nanotube and its conjugate momentum $\hat{P} = i\hbar(b^\dagger-b)/2\Delta x_{gs}$. 
This is achieved by the transformation $H\rightarrow \tilde{H}=UHU^{\dagger}$, 
where $U \equiv \exp(i\Delta x_e\hat{P}\hat{N}/\hbar)$. Here 
$\Delta x_e=
\mathfrak{F}/2M\omega^2$ is the difference 
between the equilibrium positions of the charged 
and neutral 
nanotube.  
To leading order in the small dimensionless parameters 
$\varepsilon_{t}=\Delta x_{gs}/\lambda$ and $\varepsilon_{p}= \Delta x_e/\Delta x_{gs}$ 
($\varepsilon_{p}\sim$ 0.1--0.01) the transformed tunneling Hamiltonian 
(\ref{eq:H_t}) is
%
\begin{align} \label{eq:transf_Htunn2}
\tilde{H}_t & = t_S\sum_{k,q} \left( 1-(\varepsilon_{t}+\varepsilon_{p})b + 
(\varepsilon_t-\varepsilon_{p})b^\dagger \right)c_q^\dagger a_{k,S}  \nonumber \\
{}& + t_L\sum_{k,q}\left( 1-\varepsilon_pb + \varepsilon_pb^\dagger\right) a^{\dagger}_{k,L}c_q +h.c. 
\end{align}
From Eq. (\ref{eq:transf_Htunn2}), it follows that in the Born approximation the   
rate of inelastic single-electron tunneling from the STM tip to the
nanotube accompanied by the absorption (+) or emission ($-$) of a vibron is
\begin{equation} \label{Pr}
\Gamma_{S,\pm} =\Gamma_S(\varepsilon_t^2+\varepsilon_p^2\pm 2\varepsilon_t\varepsilon_p) \,, 
\end{equation}
%
where $\Gamma_S = \Gamma_S(V, T)$ is the rate of elastic electron 
tunneling across the STM-nanotube junction.
The first (second) term of (\ref{Pr}) gives the probability for tunneling assisted by either 
absorption or emission of a vibron due to the TEM (PEM) coupling alone, while the 
third term corresponds to the ``interference'' between these two mechanisms in the
case of vibron emission ($-$) and absorption (+). Clearly, the probability for 
vibron-assisted electron tunneling is different depending on whether a vibron 
is absorbed or emitted and the difference can be controlled by the bias voltage since 
$\varepsilon_p \propto \Delta  x_e(\delta V)$.

In particular, $\Delta  x_e > 0$ if $\delta V>0$ so that the interference is destructive
(constructive) for tunneling accompanied by vibron emission (absorption). If $\delta V<0$
the situation is reversed in the sense that $\Delta  x_e < 0$ and the interference is
constructive (destructive) for emission (absorption) processes.

The case of constructive interference for emission processes has been analyzed in 
Ref. \onlinecite{Nord2005}, where it was shown that a promotion of emission
over absorption processes may lead to an electromechanical instability of the
system if $V$ exceeds a certain dissipation-dependent threshold. 
Here we will focus on the reverse situation.

A complete suppression of the emission processes would eventually drive the mechanical 
subsystem to its ground state. However, two more types of electronic transitions that may 
generate vibron emission remain to be considered. The first is the tunneling of an electron 
from the nanotube to the STM. By virtue of time reversal symmetry, the mechanism 
responsible for the suppression of vibron emission during tunneling from the STM to the 
nanotube stimulates the emission of vibrons during tunneling in the reverse direction.
In order to make the effect of such transitions negligible in the energy balance
for the mechanical subsystem, an electron that has tunneled from the STM should
escape from the nanotube to the leads before it can tunnel back to the STM by an 
inelastic transition. This requires that $|t_S| \ll |t_L|$ and $k_BT\ll eV_c$, 
where the latter constraint ensures an exponential suppression of the probability 
for electrons to tunnel from the leads to the nanotube. These conditions are satisfied 
in the real experimental situation. 

In addition to the ``backward'' tunneling transitions, vibrons can also be emitted 
when electrons tunnel from the nanotube to the leads, but then only by virtue of the 
polaronic coupling mechanism (see Eq.~(\ref{eq:transf_Htunn2})). 

From Eqs.~(\ref{Pr}) and (\ref{eq:transf_Htunn2}) and the definitions of 
$\varepsilon_t$ and $\varepsilon_p$, it follows that the ratio beween the 
total rate of vibron emission and the total rate of vibron absorption 
reaches an absolute minimum for the bias voltage $V^* = V_c-\delta V^*$ 
that verifies the condition 
\begin{align} \label{V*}
\mathfrak{F}(\delta V^*)=\frac{\hbar \omega}{\sqrt{2}\lambda} \,.
\end{align}
%


\noindent From the above considerations, we conclude that cooling of the nanotube vibrations 
can only occur for bias voltages below the Coulomb blockade threshold ($\delta V>0$). 
%
%

However, below the Coulomb blockade threshold voltage, charge transport is
blocked at zero temperature. 
The temperature needed to overcome the Coulomb blockade is determined by $k_BT \geq e\delta V$. 
On the other hand, the temperature cannot be too high, since otherwise ``backward'' 
transitions from the leads to the nanotube would no longer be negligible and possibly 
compensate for the vibrons absorbed during the ``forward'' transitions. 
These conditions restrict the range of possible temperatures to the interval: 
$eV_C\gg k_BT\geq e\delta V \cong e\delta V^*$.
The order of magnitude of the lower bound can be found by means of the 
condition 
(\ref{V*}) and by estimating the capacitance between the
STM tip and the nanotube as $C_S\simeq 2\pi \varepsilon_0 D/\ln(2h/r_0)$, where 
$h\sim 1$, $\varepsilon_0$ is the vacuum permittivity, $D\sim 10$ nm is 
the characteristic diameter of the STM tip, $r_0\sim 0.5$ nm is the nanotube radius. 
One finds that the
temperature required in order to overcome the Coulomb blockade at $\delta V^{*}$
is about 0.1~K.

For a quantitative analysis of the cooling mechanism described above, 
we followed the standard procedure to derive a generalized 
master equation for the reduced density matrix that describes the nanotube degrees 
of freedom \cite{Gorelik2005}. After tracing out the charge degrees of freedom 
and applying a perturbation approach with respect to the small parameters $\epsilon_{t,p}$
one gets a set of equations for the probabilities $p_n$ to find the nanotube in the Fock 
state $|n\rangle$ characterized by $n$ vibrons. If the rate $\Gamma_L$ of tunneling 
from the nanotube to the leads is much larger than the rate $\Gamma_S$ of tunneling 
from the STM to the nanotube these equations reduce to 
\begin{align} \label{eq:rate}
& \left[ (4n+2)\varepsilon_p^2 +(2n+1)\varepsilon_t^2-2\varepsilon_p\varepsilon_t\right]
p_n- \frac{\mathcal{L}_{\gamma} [p_{n}]}{\Gamma_S} =  \\
&\left[(\varepsilon_p+\varepsilon_t)^2+\varepsilon_p^2\right](n+1)p_{n+1}+
\left[(\varepsilon_p-\varepsilon_t)^2+\varepsilon_p^2\right]np_{n-1} \,,\nonumber
\end{align}
where $\mathcal{L}_\gamma$ describes the interaction with the environment, which takes 
the standard form \cite{Breuer2002}: $\mathcal{L}_\gamma[p_n] \equiv 
\gamma(n+1)[(n_{\rm th}+1)p_{n+1}-n_{\rm th}p_n]- \gamma n[(n_{\rm th}+1)p_n-n_{\rm th}p_{n-1}]$,
where $\gamma \equiv \omega /Q$, $Q$ being the quality factor of the nanotube resonator, 
and $n_{\rm th} = (e^{\hbar \omega/k_B T}-1)^{-1}$ is the thermal average number of vibrons. 

%
%

\begin{figure}
\center
\includegraphics[width=0.45\textwidth]{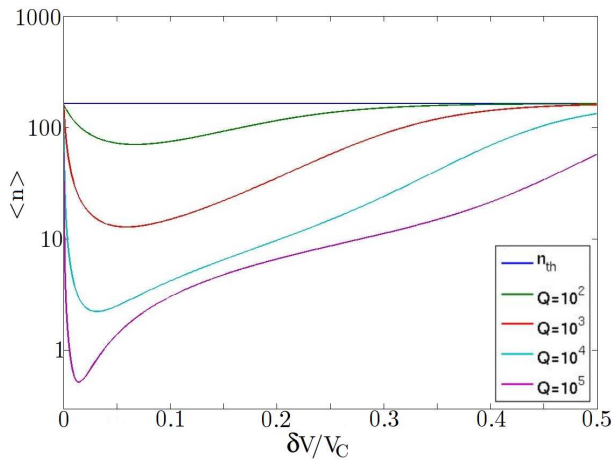}
\caption{The average number $ \langle n \rangle$ of vibrons is plotted against 
the difference $\delta V=V_C-V$ between the Coulomb blockade threshold voltage 
$V_c$ and the bias voltage $V$. Each curve corresponds to a different quality
factor of the oscillator, while the straight line gives the thermal average 
number of vibrons at the temperature of $T = 1$ K.}
\label{fig:n_ph}
\end{figure}

Equation (\ref{eq:rate}) can be solved for the stationary probability distribution 
$p_n$ with the result
\begin{align} \label{eq:stat_P}
p_n  & =  (1-r)r^n \,,  \\
& r = \frac{\varepsilon_p^{2}+(\varepsilon_t-\varepsilon_p)^2+ 
(\gamma/\Gamma_S) n_{\rm th}}
{\varepsilon_p^2+(\varepsilon_t+\varepsilon_p)^2+
(\gamma/\Gamma_S)(n_{\rm th}+1)}\nonumber
\end{align}
The average number of vibrons, $ \langle n \rangle = \sum_m m p_{m}=r/(1-r)$ is plotted
as a function of the bias voltage for several values of the quality factor in 
Fig. \ref{fig:n_ph}. The theoretical limit of the cooling efficiency
occurs for the bias voltage defined by $\varepsilon_p(\delta V^*) = \varepsilon_t/\sqrt{2}$) 
in the limit $Q \to \infty$.
Equation (\ref{eq:stat_P}) implies that the corresponding
average number of excitations is $n_{min} = (\sqrt{2}-1)/2 \approx$0.2.

In order to investigate the signatures of the cooling mechanism in a
directly measurable property, we have calculated the current $I$ 
perturbatively to second order in $\varepsilon_{t,p}$ with the result
\begin{equation} \label{eq:I}
I = I_0 \left( 1+\varepsilon_t^2 \left(1+2\langle n \rangle \right) \right)\,.
\end{equation}
Here $I_0 = e\Gamma_S(V,T)$ with 
$\Gamma_S \sim k_BT$ 
if $k_BT \gg e\delta V^*$ and $\Gamma_S$ remains independent of voltage in a 
certain voltage interval, where the differential conductance will be completely 
determined by the derivative of the average number of vibrons with respect to 
voltage, i.e. $\partial I/\partial V \cong 2 I_0\varepsilon_t^2\partial 
\langle n \rangle /\partial V$. 
Therefore, the cooling effect will be reflected in the structure of the $dI/dV-V$ 
curves and accessible for experimental investigation. 

In conclusion we have proposed a novel mechanism for ground-state cooling of 
nanomechanical resonators based on the injection of a tunneling current from a
voltage-biased STM tip. For the model system considered we have shown that 
the direction of the electrostatic force that acts on a suspended charged nanotube
can be flipped by changing the voltage bias. This makes it possible to control the
interference between two distinct contributions to the quantum mechanical probability
amplitudes for vibron absorption and emission during electron tunneling. For a bias 
voltage slightly below the Coulomb blockade threshold voltage, the probability 
amplitude for vibron emission becomes very small. At this bias a thermally activated 
current therefore leads to a cooling of the nanomechanical vibrations. 
Our analysis shows that the effective temperature that can be reached may correspond 
to an average vibron population of the fundamental bending mode as low as 0.2. 
The cooling mechanism, which should be observable by its effect on 
the differential conductance of the system, is crucially dependent on 
the Coulomb blockade phenomenon and hence on the quantization of electric charge. 

Partial support from the Swedish VR and SSF, the EC project QNEMS (FP7-ICT-233952), 
the Faculty of Science at the University of Gothenburg through its Nanoparticle Research 
Platform and from the Korean WCU program funded by MEST/NFR 
(R31-2008-000-10057-0) is gratefully acknowledged.




\end{document}